\documentclass{jpsj3}
\usepackage{txfonts}
\usepackage{color}
\usepackage{bm}

\title{Size dependent optical response in coupled systems of plasmons and {electron-hole} pairs in metallic nanostructures}

\author{Masayuki Iio$^{1}$, Tomohiro Yokoyama$^2$
\thanks{{tomohiro.yokoyama@mp.es.}osaka-u.ac.jp},
Takeshi Inaoka$^3$, and Hajime Ishihara$^{2,4}$}
\inst{$^1$Department of Physics and Electronics, Graduate School of Engineering,
Osaka Prefecture University, 1-1 Gakuen-cho, Naka-ku, Sakai, Osaka 599-8531, Japan \\
$^2$ Department of Materials Engineering Science, Graduate School of Engineering Science,
Osaka University, 1-3 Machikaneyama, Toyonaka, Osaka 560-8531, Japan\\
$^3$Department of Physics and Earth Sciences, Faculty of Science,
University of the Ryukyus, 1 Senbaru, Nishihara-cho, Okinawa 903-0213, Japan \\
$^4$Center for Quantum Information and Quantum Biology, Osaka University, 1-3 Machikaneyama, Toyonaka, Osaka 560-8531, Japan} 

\abst{
In bulk materials, the collective modes and individual modes are orthogonal each other,
and no connection occurs if there is no damping processes.
In the presence of damping, the collective modes, i.e., plasmons decay into the hot carriers.
In finite systems, {the} collective and individual modes are {coupled by the Coulomb interaction.}
Such couplings by longitudinal (L) field have been intensively investigated,
{whereas} a coupling via transverse (T) field has been poorly studied
{although the} plasmon {is} excited by {an irradiated light on surface and in finite nanostructures.
Then, the T} field {would} play a significant role in the coupling between {the} collective and individual {excitations}.
{In this study, we} investigate how {the T field} mediates {the} coherent coupling.
This study is based on the recently developed microscopic nonlocal theory of {electronic systems in metals}
and the results of eigenmode analyses by this theory.
{To tune the coupling strength in a single nanorod, we examine three parameters:
Rod length $L_z$, background refractive index $n_{\rm b}$, and Fermi energy $\varepsilon_{\rm F}$.
We discuss the modulation ratio of the spectrum of optical response coefficients to evaluate the coupling by the T field.
The T field shifts the collective excitation energy, which causes a finite modulation at both collective excitation and individual excitations.
The three parameters can change the energy distance between the collective and individual excitations.
Thus, the coherent coupling by the T field is enhanced for a proper tuning of the parameters.}
The results of the investigation of system parameter dependence would give insight into
the guiding principle of designing the materials for highly efficient hot carrier generation.
}

\begin{document}
\maketitle

\section{Introduction}
The hot-carrier generation {through {the} plasmons excited} by light has been drawing growing attention because of 
its potential applications such as photocatalysis~\cite{Ueno16}, photodetection~\cite{Li17},
photocarrier injection~\cite{Tatsuma17}, and photovoltaics~\cite{Clavero14}.
The relaxation of plasmons causes thermally non-equilibrium distribution of electrons and holes in several ten femtosecond timescale, and this distribution relaxes to a (high temperature) thermal distribution within picosecond timescale.
Such a mechanism has been studied based on phenomenological relaxation times~\cite{Brongersma15,Besteiro17}.  
This light-induced hot-carrier generation and injection~\cite{Brongersma15,Besteiro17,White12,Govorov13,Govorov14,Kumar19}
attract many attentions because of higher energy of the hot carriers than the Schottky barrier between
metals and semiconductors~\cite{Goykhman11,White12}.

The hot-carrier generation depends on the size and shape of the metallic nanostructure hosting the plasmons~\cite{Govorov13,Govorov14} because, in nanostructures,  
quantum coherence of confined electrons becomes significant.
Therefore, the nanotechnologies to fabricate precisely controlled nanostructures are crucial  
for the plasmonics.
For understanding the mechanism laying behind the phenenomena and developing the
carrier generation devices, the study of the interplay between 
the collective plasmon excitation and the individual electron-hole pair excitations 
from the microscopic view point is a significant subject.
Further, the design of sample structures are crucial also for controlling 
the electromagnetic (EM) fields to increase an efficiency of hot-carrier generation and injection~\cite{XShi18,YELiu23}.
Therefore, the study of plasmonics based on the self-consistent manner for the EM field and
plasmons modulated by the sample structures is significant for developing efficient plasmonic devices.

In the bulk of metallic samples, the collective excitation and the individual excitations 
are orthogonal to each other, and they are not connected if the damping processes are absent.
However, if the translational symmetry of the system is broken, these two excitation modes have
{an} interplay and form hybridized modes.
{S}uch interplay has been studied as a coherent coupling via the longitudinal (L) component of the electric field based on {the} 
first-principle calculation for nanoscale clusters~\cite{Ma15,You18}.
{U}nder the broken translational symmetry, plasmons interact with transverse (T) component of the EM field,
and {the} T field should be described self-consistently with the collective and the individual excitations.
As one aspect of interaction between
{the} plasmons and {the} T field has been discussed as the nonlocal response
based on the hydrodynamic model that relates the electric field and the spatial gradient of current density in addition to a usual Drude conductivity, which describes phenomenologically a nonlocal response and supplies an {applicable} method to numerical calculation~\cite{Bennett70,Schwartz82,Pitarke07,Mortensen14,Christensen14,Svendsen20}. 

{The} T field should have significant roles to mediate the collective and the individual excitations under some conditions. 
In our recent study~\cite{Yokoyama22}, we have formulated the microscopic nonlocal theory and
pointed out that the T field gives a finite contribution to the coupling between the collective and the individual excitations
showing the numerical demonstration by using small bases of electronic systems.
Further, we have investigated the sample size that this effect starts to appear by demonstrating
the dependence of T field-mediated shift of {the} collective excitation energy on
{the system parameters~\cite{iio1}.}
In the formulation, the constitutive equation with the nonlocal susceptibility and the Maxwell's equations
are considered self-consistently~\cite{book:cho1}.
The susceptibility is calculated from the microscopic Hamiltonian and the linear response theory~\cite{Kubo57}.
Because of the spatial extent of the confined electronic wavefunctions, the nonlocality 
depending on the system size and shape appears in the optical response.
By using a single nanorod, we have analyzed the system eigenmodes.
{Owing to the L component of EM field, one isolated collective excitation and continuously distributed individual excitaions are formed.
When the T field contribution is considered, we have} found a finite shift of collective excitation energy
due to the {T field-mediated} coherent coupling between the collective and the individual excitations~\cite{Yokoyama22,iio1}.

In this study, we investigate {the T field}-mediate coherent coupling between the collective and the individual excitations
that appears in optical response. Considering the same model as in Ref.\ \cite{iio1},
i.e., a nanoscale nanorod, we see spectra of {the} induced current density under light irradiation in a certain condition.
By switching the T field in the calculation of Green's function representing interaction among {electron-hole pair states,}
we examine how the effect of {T field}-mediate coherent coupling appears in the spectra of {the} induced current
arising from the transitions due to the collective and individual excitations.
The result indicates a significant appearance of the {T field}-mediate coherent coupling, which 
suggests the possibility of spontaneous resonance between localized surface plasmon polaritons
and electron-hole pairs excitations.  
Thus, this study would lead to a guiding principle to improve the efficiency of {hot carrier} generation
by the localized surface plasmon resonance.

The remainder of this article is structured as follows.
In Sec.\ \ref{sec:theory}, we describe the microscopic Hamiltonian and a self-consistent formulation for
the nonlocal linear response and the Maxwell's equations.
{The model system is described in Sec.\ \ref{sec:model}. 
The formulation is applied to numerical calculations for nanorods.
In Sec.\ \ref{sec:excitation:spectra}, we show calculated spectra of induced current densities
{with tuning of several system parameters} and discuss the {effect by} T field appearing in
the shift of collective excitation and {the} coherent coupling between {the collective and individual} excitations. 
Sec.\ \ref{sec:conculusion} is devoted to the summary and conclusions.}

\section{Theoretical framework}
\label{sec:theory}
In this study, we use the theoretical method developed in our previous study~\cite{Yokoyama22},
where the self-consistent treatment of the constitutive and Maxwell's equations provides a matrix equation.
The solutions of this equation provide amplitudes of respective components of 
induced current density when they are expanded with the system eigenmodes. 
The details of the theory are described in our previous paper~\cite{Yokoyama22}. 

\subsection{Hamiltonian for the light--matter interaction}
The electromagnetic fields interact with the
{electrons in materials. The Hamiltonian of the interaction between the electrons and the fields}
can be expressed as follows:
\begin{eqnarray}
\hat{H} &=& \sum_j
\frac{\left\{ \hat{\bm{p}}_j - e \int d{\bm{x}} \bm{A} ({\bm{x}},t) \delta ({\bm{x} - \hat{\bm{x}}_j}) \right\}^2}
{2{m^*}}
\nonumber \\
& & + e \sum_j  \int d\bm{x} \phi (\bm{x},t) \delta (\bm{x} - \hat{\bm{x}}_j)
\nonumber \\
& & + \frac{1}{2} \sum_{i \ne j} \frac{e^2}{4\pi \epsilon_0 |\hat{\bm{x}}_i - \hat{\bm{x}}_j|}
{.}
\label{eq:singleH}
\end{eqnarray}
In this equation,
{$m^*$}
represents the electron mass, $e$ is the electron charge, and $\epsilon_0$ is the vacuum dielectric constant.
{We employ t}he Coulomb gauge
where the vector potential {$\bm{A} (\bm{x},t)$} satisfies $\text{div} \bm{A} = 0$ and represents the T field.
The scalar potential $\phi (\bm{x},t)$ contributes to the L field, and it arises from both the nuclei and external sources,
i.e., $\phi (\bm{x},t) = \phi_{\rm ncl} (\bm{x}) + \phi_{\rm ext} (\bm{x},t)$.
The {Coulomb} interaction between electrons, which is part of the L-component, is also taken into account in the third term of the Hamiltonian.

By applying the concept of second quantization and utilizing a mean-field approximation (refer to Ref.\ \cite{Yokoyama22} for details), the Hamiltonian can be decomposed as follows:
\begin{equation}
\hat{H} = \hat{H}_0 + \hat{H}^\prime {.}
\label{eq:fullH}
\end{equation}
Here, $\hat{H}_0$ represents
{the Hamiltonian without the external L and T fields by $\bm{A}$ and $\phi_{\rm ext}$, respectively:
\begin{eqnarray}
\hat{H}_0 &=& \sum_{n,n^\prime} \psi_{n^\prime}^* (\bm{x})
\left( -\frac{\hbar^2}{2m^*} \bm{\nabla}_x^2 \right)
\psi_n (\bm{x}) \hat{a}_{n^\prime}^\dagger \hat{a}_n
\nonumber \\
& & + \int d\bm{x} \sum_{n,n^\prime} \hat{a}_{n^\prime}^\dagger \psi_{n^\prime}^* (\bm{x})
\left[ \phi_{\rm ncl} (\bm{x}) + \hat{\phi}_{\rm e-e} (\bm{x}) \right] \psi_n (\bm{x}) \hat{a}_n
\label{eq:H0}
\end{eqnarray}
with the part of scalar potential owing to the electron-electron interaction,
\begin{equation}
\hat{\phi}_{\rm e-e} (\bm{x}) = \int d\bm{x}^\prime
e \sum_{m, m^\prime} \frac{\psi_{m^\prime}^* (\bm{x}^\prime) \psi_m (\bm{x}^\prime)}{4\pi \epsilon_0 |\bm{x} - \bm{x}^\prime|}
 \hat{a}_{m^\prime}^\dagger \hat{a}_m.
\label{eq:eescalar}
\end{equation}
Here, $\hat{a}_n^\dagger$ and $\hat{a}_n$ are the creation and annihilation operators of electron of the state $n$, respectively.
$\psi_n (\bm{x})$ is the wavefunction of the state $n$.
}

{Under the external fields, we obtain the perturbative Hamiltonian $H^\prime$.
We consider a monochromatic electromagnetic fields, $\bm{A} (\bm{x},t) = \bm{A} (\bm{x};\omega) e^{-i\omega t}$ and
$\phi_{\rm ext} (\bm{x},t) = \phi_{\rm ext} (\bm{x};\omega) e^{-i\omega t}$.
Then, the perturbative Hamiltonian becomes
\begin{equation}
\hat{H}^\prime \simeq - \int d\bm{x} \left[
\hat{\bm{j}} (\bm{x}) \cdot \bm{A} (\bm{x};\omega) - \delta \hat{\rho} (\bm{x}) \hat{\phi} (\bm{x};\omega)
\right] e^{-i\omega t}
\label{eq:Hprime}
\end{equation}
with
the induced current and the (deviation of) charge density operators,
\begin{eqnarray}
\hat{\bm{j}} (\bm{x}) &=& \sum_{n,n^\prime} \bm{j}_{n^\prime n} (\bm{x}) \hat{a}_{n^\prime}^\dagger \hat{a}_n, \\
\delta \hat{\rho} (\bm{x}) &=& \sum_{n,n^\prime} \rho_{n^\prime n} (\bm{x}) \hat{a}_{n^\prime}^\dagger \hat{a}_n - \rho_0 (\bm{x}).
\end{eqnarray}
Note that $\rho_0 (\bm{x}) = e \sum_n \psi_n^* (\bm{x}) \psi_n (\bm{x}) \langle \hat{a}_n^\dagger \hat{a}_n \rangle_0$ is
the charge density in the static situation, hence $\delta \hat{\rho} (\bm{x})$ describes
negative and positive charge density due to the electron excitaion and the nuclei.
The matrix elements for the current and charge are}
\begin{eqnarray}
\bm{j}_{n^\prime n} (\bm{x}) &=& - \frac{e\hbar}{2im^*} \Big[ \left\{ \bm{\nabla}_x \psi_{n^\prime}^* (\bm{x}) \right\} \psi_n (\bm{x})
- \psi_{n^\prime}^* (\bm{x}) \left\{ \bm{\nabla}_x \psi_n (\bm{x}) \right\} \Big] \nonumber \\
& & - \frac{e^2}{m^*} \psi_{n^\prime}^* (\bm{x}) \bm{A} (\bm{x};\omega) \psi_n (\bm{x}), \\
\rho_{n^\prime n} (\bm{x}) &=& e \psi_{n^\prime}^* (\bm{x}) \psi_n (\bm{x}).
\end{eqnarray}
{In the second tern in Eq.\ (\ref{eq:Hprime}), the scalar potential describes the external L field and
additional field by the induced polarized charge,
\begin{equation}
\hat{\phi} (\bm{x};\omega) = \phi_{\rm ext} (\bm{x};\omega) + \hat{\phi}_{\rm pol} (\bm{x})
\label{eq:totalphi}
\end{equation}
with
}
\begin{equation}
\hat{\phi}_{\rm pol} (\bm{x}) = \int d\bm{x}^\prime
\frac{\delta \hat{\rho} (\bm{x}^\prime)}{4\pi \epsilon_0 |\bm{x} - \bm{x}^\prime|}.
\label{eq:polscalar}
\end{equation}
{It is worthy to note that $\delta \hat{\rho} (\bm{x}) \hat{\phi}_{\rm pol} (\bm{x})$
originates from the Coulomb interaction.
In the absence of external fields, this term vanishes, and the Coulomb interaction is taken into account only in $H_0$ in Eq.\ (\ref{eq:H0}).
In this term, not only electron-electron interaction but also electron-nuclei interaction are included.}

\subsection{{Self-consistent equation in matrix form}}
From the Hamiltonian $\hat{H} = \hat{H}_0 + \hat{H}^{\prime}$, a nonlocal susceptibility can be derived.
{For the deviation of susceptibility $\bar{\mathcal{X}} (\bm{x},\bm{x}^{\prime} ;\omega)$, we take the statistical average for
$\hat{\bm{j}}$, $\delta \hat{\rho}$, and $\hat{\phi}$ under the mean field approximation~\cite{Yokoyama22}.}
Utilizing a four-vector representation
{
\begin{equation}
\bm{\mathcal{A}} (\bm{x}{;\omega})
= \left( \begin{matrix} \bm{A} (\bm{x}{;\omega}) \\
-\frac{\phi (\bm{x}{;\omega}) }{c} \end{matrix} \right)
\end{equation}
for the vector and scalar potentials describing the Maxwell's fields and
\begin{equation}
\bm{\mathcal{J}} (\bm{x}{;\omega})
= \left( \begin{matrix} \bm{j} (\bm{x}{;\omega}) \\
c\delta \rho (\bm{x}{;\omega})  \end{matrix} \right)
\end{equation}
for the current and charge densties describing the electronic response,
}
the constitutive equation can be expressed as:
\begin{equation}
\bm{\mathcal{J}} (\bm{x};\omega) = \bm{\mathcal{J}}_0 (\bm{x};\omega) + \int d\bm{x}^{\prime} \bar{\mathcal{X}} (\bm{x},\bm{x}^{\prime} ;\omega) \bm{\mathcal{A}} (\bm{x}^{\prime} ;\omega).
\label{eq:cons4vec}
\end{equation}
The first term $\bm{\mathcal{J}}_0 {(\bm{x};\omega)}$ corresponds to the contribution from the average {current} density. The nonlocal susceptibility is described using matrix elements $\bm{\mathcal{J}}_{\nu \mu} = \langle \nu | \hat{\bm{\mathcal{J}}} | \mu \rangle$ of the densities, leading to the expression:
\begin{equation}
\bar{\mathcal{X}} (\bm{x},\bm{x}^{\prime} ;\omega) = \sum_{\mu ,\nu} \left[ f_{\nu \mu} (\omega) \bm{\mathcal{J}}_{\mu \nu} (\bm{x}) \left( \bm{\mathcal{J}}_{\nu \mu} (\bm{x}^{\prime}) \right)^{\rm t} + h_{\nu \mu} (\omega) \bm{\mathcal{J}}_{\nu \mu} (\bm{x}) \left( \bm{\mathcal{J}}_{\mu \nu} (\bm{x}^{\prime}) \right)^{\rm t} \right].
\label{eq:nonlocalsus4vec_modified}
\end{equation}
In these equations, $|\mu \rangle$ represents the eigenstates of $\hat{H}_0$, and the factors 
$f_{\nu \mu} (\omega) = \eta_\mu/(\hbar \omega_{\nu \mu} - \hbar \omega {- i\hbar \gamma})$ 
and 
$h_{\mu \nu} (\omega) = \eta_\mu/(\hbar \omega_{\nu \mu} + \hbar \omega {+ i\hbar \gamma})$
are defined with respect to the energy differences $\hbar \omega_{\nu \mu} = E_\nu - E_\mu$ considering causality through the imaginary infinitesimal value $\gamma$. The numerator terms involve
$\eta_{\mu} = \langle \mu |e^{-\beta \hat{H}_0}| \mu \rangle /Z_0$ with $Z_0 = {\rm Tr} \{ e^{-\beta \hat{H}_0} \}$.
{At zero temperature, this factor becomes $\eta_{\mu} = \delta_{0\mu}$. In the following, we assume zero temperature $T=0$.}

Induced currents and charge densities $\bm{\mathcal{J}} {(\bm{x};\omega)}$ act as sources for the response fields in Maxwell's equations. {For the four-vector} representation, {the} Maxwell's equations {for the vector and scalar potentials} are summarized as
\begin{equation}
\bar{\mathcal{D}}(\bm{x};\omega) \bm{\mathcal{A}}(\bm{x};\omega) = -\mu_0 \bm{\mathcal{J}}(\bm{x};\omega) \label{eq:Maxwell4vec}
\end{equation}
with
\begin{equation}
\bar{\mathcal{D}}(\bm{x};\omega) =
\begin{pmatrix}
\bm{\nabla}^2 + \left(\frac{\omega}{c}\right)^2 & -i\left(\frac{\omega}{c}\right)\bm{\nabla} \\
0 & -\bm{\nabla}^2
\end{pmatrix}.
\end{equation}
The formal solution of Eq. \eqref{eq:Maxwell4vec} is
\begin{equation}
\bm{\mathcal{A}}(\bm{x};\omega) = \bm{\mathcal{A}}_0(\bm{x};\omega) - \mu_0 \int d\bm{x}' \bar{\mathcal{G}}(\bm{x},\bm{x}';\omega) \bm{\mathcal{J}}(\bm{x}';\omega) \label{eq:formalA4vec}
\end{equation}
where the Green's function satisfies
\begin{equation}
\bar{\mathcal{D}}(\bm{x};\omega) \bar{\mathcal{G}}(\bm{x},\bm{x}';\omega) = \delta(\bm{x} - \bm{x}'). \label{eq:Green4vec}
\end{equation}
The first term represents an ``incident field,'' as $\bar{\mathcal{D}}(\bm{x};\omega) \bm{\mathcal{A}}_0(\bm{x};\omega) = 0$.

The constitutive equation \eqref{eq:cons4vec} and the solution to Maxwell's equations \eqref{eq:formalA4vec} are in a self-consistent relationship. This leads to a self-consistent equation. To make it solvable, we {apply Eq.\ (\ref{eq:cons4vec}) into Eq.\ (\ref{eq:formalA4vec}),
\begin{equation}
\bm{\mathcal{A}}(\bm{x}) = \bm{\mathcal{A}}_0(\bm{x}) - \mu_0 \int d\bm{x}' \bar{\mathcal{G}}(\bm{x},\bm{x}') \bm{\mathcal{J}}_0(\bm{x}')
- \mu_0 \int d\bm{x}'d\bm{x}'' \bar{\mathcal{G}}(\bm{x},\bm{x}') \bar{\mathcal{X}} (\bm{x}',\bm{x}'') \bm{\mathcal{A}} (\bm{x}''),
\end{equation}
and} multiply $\left[ \bm{\mathcal{J}}_{\nu' 0} (\bm{x}) \right]^{\rm t}$ {and $\left[ \bm{\mathcal{J}}_{0 \nu'} (\bm{x}) \right]^{\rm t}$} from the left and integrate with respect to $\bm{x}$. {Here, we reduce $\omega$ for simplification. Then, we obtain
\begin{eqnarray}
\left(\hbar \omega_{\nu' 0} - \hbar \omega - i\gamma \right) X_{\nu'}^{(-)} &=& Y_{\nu'}^{(-)}
+ \sum_{j,\alpha} U_{\nu' j \alpha} X_{j\alpha}^{(A)} - \sum_{\nu} \left[ K_{\nu' \nu} X_{\nu}^{(-)} + L_{\nu' \nu} X_{\nu}^{(+)} \right], \label{eq:exSCmatrix1} \\
\left(\hbar \omega_{\nu' 0} + \hbar \omega + i\gamma \right) X_{\nu'}^{(+)} &=& Y_{\nu'}^{(+)}
+ \sum_{j,\alpha} V_{\nu' j \alpha} X_{j\alpha}^{(A)} - \sum_{\nu} \left[ M_{\nu' \nu} X_{\nu}^{(-)} + N_{\nu' \nu} X_{\nu}^{(+)} \right] \label{eq:exSCmatrix2}
\end{eqnarray}
}
with
\begin{eqnarray}
X_{\nu}^{(-)} &=& \frac{1}{\hbar \omega_{\nu 0} - \hbar \omega {- i\hbar \gamma}} \int d\bm{x} \left[\bm{\mathcal{J}}_{\nu 0}(\bm{x})\right]^{t} \bm{\mathcal{A}}(\bm{x}), \label{eq:X-JA} \\
X_{\nu}^{(+)} &=& \frac{1}{\hbar \omega_{\nu 0} + \hbar \omega {+ i\hbar \gamma}} \int d\bm{x} \left[\bm{\mathcal{J}}_{0 \nu}(\bm{x})\right]^{t} \bm{\mathcal{A}}(\bm{x}), \label{eq:X+JA} \\
X_{j \alpha}^{(A)} &=& \int d\bm{x} \left[ \varphi_j(\bm{x}) \bm{e}_\alpha \right]^{t} \bm{\mathcal{A}}(\bm{x}), \label{eq:exSFXA} \\
Y^{(-)}_{\nu} &=& \int d\bm{x} \left[\bm{\mathcal{J}}_{\nu 0}(\bm{x})\right]^{t} \bm{\mathcal{A}}_0(\bm{x}), \label{eq:Y-JA} \\
Y^{(+)}_{\nu} &=& \int d\bm{x} \left[\bm{\mathcal{J}}_{0 \nu}(\bm{x})\right]^{t} \bm{\mathcal{A}}_0(\bm{x}), \label{eq:Y+JA}
\end{eqnarray}
and
{
\begin{eqnarray}
K_{\nu' \nu} &=& \mu_0  \int d\bm{x} d\bm{x}' \left[\bm{\mathcal{J}}_{\nu' 0}(\bm{x})\right]^{t} \bar{\mathcal{G}}(\bm{x},\bm{x}') \bm{\mathcal{J}}_{0 \nu}(\bm{x}'), \label{eq:KJGJ} \\
L_{\nu' \nu} &=& \mu_0  \int d\bm{x} d\bm{x}' \left[\bm{\mathcal{J}}_{\nu' 0}(\bm{x})\right]^{t} \bar{\mathcal{G}}(\bm{x},\bm{x}') \bm{\mathcal{J}}_{\nu 0}(\bm{x}'), \label{eq:LJGJ} \\
M_{\nu' \nu} &=& \mu_0  \int d\bm{x} d\bm{x}' \left[\bm{\mathcal{J}}_{0 \nu'}(\bm{x})\right]^{t} \bar{\mathcal{G}}(\bm{x},\bm{x}') \bm{\mathcal{J}}_{0 \nu}(\bm{x}'), \label{eq:MJGJ} \\
N_{\nu' \nu} &=& \mu_0  \int d\bm{x} d\bm{x}' \left[\bm{\mathcal{J}}_{0 \nu'}(\bm{x})\right]^{t} \bar{\mathcal{G}}(\bm{x},\bm{x}') \bm{\mathcal{J}}_{\nu 0}(\bm{x}'), \label{eq:NJGJ} \\
U_{\nu' j \alpha} &=& \left( \frac{\omega_{\rm p}}{c} \right)^2 \int d\bm{x} d\bm{x}' \left[\bm{\mathcal{J}}_{\nu' 0}(\bm{x})\right]^{t} \bar{\mathcal{G}}(\bm{x},\bm{x}') \left[ \varphi_j^*(\bm{x}') \bm{e}_\alpha \right], \label{eq:exSFU} \\
V_{\nu' j \alpha} &=& \left( \frac{\omega_{\rm p}}{c} \right)^2 \int d\bm{x} d\bm{x}' \left[\bm{\mathcal{J}}_{0\nu'}(\bm{x})\right]^{t} \bar{\mathcal{G}}(\bm{x},\bm{x}') \left[ \varphi_j^*(\bm{x}') \bm{e}_\alpha \right]. \label{eq:exSFV}
\end{eqnarray}
The coefficients $X^{(\pm)}_{\nu}$ and $X^{(A)}_{j\alpha}$ are for the respond fields and $Y^{(\pm)}_{\nu}$ is for the incident fields.}
For the $\bm{\mathcal{J}}_0 {(\bm{x})}$ term,
we apply $\delta(\bm{x}-\bm{x}') = \sum_m \varphi_m^*(\bm{x}) \varphi_m(\bm{x}')$ and use the plasma frequency in
{three-dimensional} bulk, $\omega_{\rm p} = \sqrt{e^2 n_0/(\epsilon_0 m^*)}$
{with $n_0 = (2m^* \varepsilon_{\rm F}/\hbar^2)^{3/2} / (3\pi^2)$ being the density of electrons}.
{For Eqs.\ (\ref{eq:exSCmatrix1}) and (\ref{eq:exSCmatrix2}),
the factors $X_{\nu}^{(\pm)}$, $X_{j \alpha}^{(A)}$, and $Y^{(\pm)}_{\nu'}$ are
treated as vectors $\bm{X}^{(\pm,A)}$ and $\bm{Y}^{(\pm)}$, respectively.
$K_{\nu' \nu}$, $L_{\nu' \nu}$, $M_{\nu' \nu}$, $N_{\nu' \nu}$, $U_{\nu' j \alpha}$, and $V_{\nu' j \alpha}$ form a matrix:
\begin{equation}
\left[\begin{matrix} \hbar \bar{\Omega} - (\hbar \omega + i\gamma)\bar{1} & \\
                            & \hbar \bar{\Omega} + (\hbar \omega + i\gamma)\bar{1} \end{matrix} \right]
\left( \begin{matrix} \bm{X}^{(-)} \\ \bm{X}^{(+)} \end{matrix} \right)
= \left( \begin{matrix} \bm{Y}^{(-)} \\ \bm{Y}^{(+)} \end{matrix} \right)
+ \left[\begin{matrix} \bar{U} & \\ & \bar{V} \end{matrix} \right]
\left( \begin{matrix} \bm{X}^{(A)} \\ \bm{X}^{(A)} \end{matrix} \right)
- \left[\begin{matrix} \bar{K} & \bar{L} \\ \bar{M} & \bar{N} \end{matrix} \right]
\left( \begin{matrix} \bm{X}^{(-)} \\ \bm{X}^{(+)} \end{matrix} \right)
\label{eq:SFmatrix0}
\end{equation}
}
By multiplying $[\varphi_{m'}(\bm{x}) \bm{e}_\beta^{\rm t}]$, we obtain aother equation.
{By combining Eq.\ (\ref{eq:SFmatrix0}) and that eqaution,}
we obtain the matrix form of the self-consistent equation
{for the respond and incident fields,
$\bm{X} = (\bm{X}^{(-)} , \bm{X}^{(+)})^{\rm t}$ and $\bm{Y} = (\bm{Y}^{(-)} , \bm{Y}^{(+)})^{\rm t}$,}
\begin{equation}
\bar{\Xi} \bm{X} = \bm{Y}.
\label{eq:exSCmat}
\end{equation}
{Here, a matrix $\bar{\Xi} (\omega)$ is constracted by the matrices in Eq.\ (\ref{eq:SFmatrix0}) (see Ref.\ \cite{Yokoyama22} for detail).}
$\bar{\Xi} (\omega)$ represents the spectrum of individual and collective excitations of electrons and holes.
{Hence, the electronic spectrum is evaluated from the matrix $\bar{\Xi} (\omega)$.
The repond fields $\bm{\mathcal{A}}(\bm{x}{;\omega})$ are discussed
in the terms of the components of induced both densities $X_{\nu}^{(\pm)} (\omega)$ in the present formulation.
The components are calculated by
\begin{equation}
\bm{X} = \bar{\Xi}^{-1} \bm{Y}
\label{eq:exSCmat2}
\end{equation}
in following sections. Via the matrix and vector structures of $\bar{\Xi}^{-1}$ and $\bm{Y}$,
a spatial property of the repond field $\bm{X}$ depending on the angle of incident field and the shape and size of nanostructure is obtained.}

The evaluation of the matrix elements for $\bar{\Xi}(\omega)$ is implemented through numerical calculations.
In {Eqs.\ \eqref{eq:KJGJ}--\eqref{eq:exSFV}}, multiple real space integrals need to be computed.
However, by performing a Fourier transformation for the current and charge densities and for the Green's function,
the number of integrals can be reduced.~\cite{Yokoyama22}
Therefore, we examine the numerical calculations in the Fourier-transformed space.
{
The Green's function is represented as
\begin{equation}
\bar{\mathcal{G}}_{\bm{k}} (\omega) =
\left( \begin{matrix}
\frac{1}{-\bm{k}^2 + (\omega/c)^2} \bar{1} &
-\frac{1}{\bm{k}^2} \frac{1}{-\bm{k}^2 + (\omega/c)^2} \frac{\omega}{c} \bm{k} \\
0 &
- \frac{1}{\bm{k}^2}
\end{matrix} \right).
\label{eq:solGreenMat_k}
\end{equation}
The components describe the current-current, current-charge, and charge-charge (Coulomb) interactions.
In the factors $K_{\nu' \nu}$, $L_{\nu' \nu}$, $M_{\nu' \nu}$, and $N_{\nu' \nu}$,
the three components in $\bar{\mathcal{G}}_{\bm{k}}$ describes the T-, T-L hybridization, and L-components of the electromagnetic fields.
By introducing a switching parameter to the first and second components,
we can caonsider both cases of the presence and absence of T-component to discuss the coherent coupling by the T field.
}
In Section \ref{sec:excitation:spectra}, we investigate how this value is modified by
T field-mediated coherent coupling between {the} collective and individual excitations.

\section{Model}
\label{sec:model}

Here, we introduce the model for the examination of spectra of the induced current densities.
We assume the situation that a metallic nanorod confining electrons is irradiated with the plane wave 
monochromatic light.

\subsection{Rectangular nanorods}
Electron and hole wavefunctions in a rectangular nanorod are given as
\begin{eqnarray}
\psi_{({\rm e} {\nu}) = (n_x,n_y,n_z)} (\bm{x})
&=& \sqrt{\frac{2}{L_x}} \sin \left( \frac{n_x \pi}{L_x} x \right)
    \sqrt{\frac{2}{L_y}} \sin \left( \frac{n_y \pi}{L_y} y \right)
    \sqrt{\frac{2}{L_z}} \sin \left( \frac{n_z \pi}{L_z} z \right),
\label{eq:wavefunc_e} \\
\psi_{({\rm h}\bar{{\nu}}) = (\bar{n}_x,\bar{n}_y,\bar{n}_z)} (\bm{x})
&=& \sqrt{\frac{2}{L_x}} \sin \left( \frac{\bar{n}_x \pi}{L_x} x \right)
    \sqrt{\frac{2}{L_y}} \sin \left( \frac{\bar{n}_y \pi}{L_y} y \right)
    \sqrt{\frac{2}{L_z}} \sin \left( \frac{\bar{n}_z \pi}{L_z} z \right)
\label{eq:wavefunc_h}
\end{eqnarray}
with $L_x$, $L_y$, and $L_z$ being the length of nanorod in the $x$, $y$, and $z$ directions, respectively.
We suppose that the basis $|{\nu} =(\rm{e}{\nu} ,\rm{h}\bar{{\nu}}) \rangle$ for the Hamiltonian $\hat{H}_0$ are given by
Eqs.\ (\ref{eq:wavefunc_e}) and (\ref{eq:wavefunc_h}), which enables us to discuss a relation between the 
appearance of coherent coupling and the sample structures. 
In this study, the parameter tuning of nanorods is significant.
{For the Fermi energy $\varepsilon_{\rm F}$ of the nanorods,} the state $|{\nu} =(\rm{e}{\nu} ,\rm{h}\bar{{\nu}}) \rangle$
{consisting of the electron and hole energies, $\varepsilon_{{\rm e}{\nu}}$ and $\varepsilon_{{\rm h} \bar{{\nu}}}$ satisfies}
$\varepsilon_{{\rm e}{\nu}} > \varepsilon_{\rm F} \ge \varepsilon_{{\rm h} \bar{{\nu}}}$.

\subsection{Incident light}
{
We consider a plane wave incident.
For the Coulomb gauge, the transverse and longitudinal fields are described by the vector and scalar potentials, respectively:
$\bm{E}^{\rm (T)} (\bm{x},t) = -\frac{\partial}{\partial t} \bm{A} (\bm{x},t) = i\omega \bm{A} (\bm{x};\omega) e^{-i\omega t}$
and $\bm{E}^{\rm (L)} (\bm{x},t) = -\bm{\nabla} \phi (\bm{x};\omega) e^{-i\omega t}$.
The incident wave consists only of transverse field. Hence we consider the incident wave as
\begin{equation}
\bm{\mathcal{A}}_0 (\bm{x};\omega)
=  \left( \begin{matrix} A_0 \bm{n}_{\rm inc} \\ 0 \end{matrix} \right) e^{i \bm{k}_{\rm inc} \cdot \bm{x}}
\label{eq:4A_0k}
\end{equation}
with a unit vector
$\bm{n}_{\rm inc} = (\cos \varphi_{\rm inc} \sin \theta_{\rm inc}, \sin \varphi_{\rm inc} \sin \theta_{\rm inc}, \cos \theta_{\rm inc})^{\rm t}$.
Here, $\theta_{\rm inc}$ and $\varphi_{\rm inc}$ are the polar and azimuthal angles with respect to the $z$-axis.
$\bm{k}_{\rm inc}$ is a wave vector of the incident wave. From the Coulomb gauge condition,
the wave vector should be $\bm{k}_{\rm inc} \cdot \bm{n}_{\rm inc} =0$.
}

\subsection{{Considering system}}

We calculate the eigenvalues of the matrix $\bar{\Xi}$ to describe the excitation spectrum.
The coherent coupling between plasmons and carriers strongly {depends} on background dielectric and spatial structures of sample.
Thus, we examine modulation of the nanorod length $L_z$ to tune the spatial correlation,
whereas the nanorod thickness is fixed at $L_x = 10\, \mathrm{nm}$ and $L_y = 15\, \mathrm{nm}$
to hold the subband structures by the confinement.
The Fermi energy of conduction electron is
{set at} $\varepsilon_{\rm F} = 3.0$ or 5.0 $\mathrm{eV}$.
The effective mass is {$m^* = 0.07 m_{\rm e}$} with $m_{\rm e}$ being the electron mass in vacuum~\cite{com1}.
For a typical size scale $L_0 = 100\, \mathrm{nm}$, an order of the confinement energy is
$E_0 = \hbar^2 \pi^2 / (2 m^* L_0^2) \simeq {5.37 \times 10^{-4}}\, \mathrm{eV}$.
We put $\gamma = 0.1 \times E_0$ as an infinitesimal value.
We focus on the excitation with $n_x = \bar{n}_x$, $n_y = \bar{n}_y$, and $n_z = \bar{n}_z +1$ to consider
the plasmon spectrum at a small wavenumber
$|\bm{q}| =\pi \sqrt{(\frac{n_x - \bar{n}_x}{L_x})^2 + (\frac{n_y - \bar{n}_y}{L_y})^2 + (\frac{n_z - \bar{n}_z}{L_z})^2} = \pi /L_z$.
{For our consideration at $|\bm{q}| = \pi /L_z$ with $\varepsilon_{\rm F} = 3.0\, \mathrm{eV}$ and $L_z = 500\, \mathrm{nm}$,
the number of bases of the bare individual excitaions, namely, electron-hole pair states given by $| \nu \rangle$ is $N=114$
including the spin degrees of freedom, and $N$ increases with an increase of $\varepsilon_{\rm F}$.}
Note that the above assumed electronic system plays sufficient role for the present purpose to reveal
the possibility of T field-mediated coupling though it does not represent full electronic systems in considered sample scales.

Because of the strong confinement in the $x$- and $y$-directions, the contributions of incident light is maximum
when the incident light's wave vector is perpendicular to the $z$-axis and
{the} vector potential is polarized only in the $z$-direction ($\theta_k=\frac{\pi}{2}$).
{We set $\bm{n}_{\rm inc} = (0,0,1)^{\rm t}$.}

{
To discuss an effect of the wavelength of transverse fields on the coherent coupling,
we examine a tuning of background refractive index $n_{\rm b} = \sqrt{\epsilon_{\rm b}/\epsilon_0}$ of the nanostructure and enviroment.
Then, the matrix components are modulated as, e.g.,
\begin{eqnarray}
K_{\nu' \nu} &=& \mu_0 \int d\bm{k} \left\{ \zeta^2 \left(\tilde{\bm{j}}_{\nu' 0}(-\bm{k})\right)^{t} \left(\frac{1}{-\bm{k}^2 + (n_{\rm b} \omega/c)^2}\right)
\tilde{\bm{j}}_{0 \nu}(\bm{k}) \right. \nonumber \\
& & + \zeta \left(\tilde{\bm{j}}_{\nu' 0}(-\bm{k})\right)^{t} \left(-\frac{1}{\bm{k}^2}\right) \left(\frac{n_{\rm b} \omega/c}{-\bm{k}^2 + (n_{\rm b} \omega/c)^2}\right) \bm{k} \left(c/n_{\rm b}\right) \tilde{\rho}_{0 \nu}(\bm{k}) \nonumber \\
& & \left. + \left(c/n_{\rm b}\right) \tilde{\rho}_{\nu' 0}(-\bm{k}) \left(\frac{1}{\bm{k}^2}\right) \left(c/n_{\rm b}\right) \tilde{\rho}_{0 \nu}(\bm{k}) \right\}. \label{eq:Knunu}
\end{eqnarray}
Here, $n_{\rm b} \omega/c$ represents the wavenumber of light in the nanorod and environment with $n_{\rm b}$.
In the first term in Eq.\ (\ref{eq:Knunu}), the transverse field mediates the interaction between
the current densities with the modulated wavenumber in the denominator.
In addition to the wavenumber modulation, the third term indicates a screening effect of the Coulomb interaction
between the charge densities by $n_{\rm b}$.
Hence, the effect of the background refractive index leads to a complicated modulation of the coherent coupling.
In Eq.\ (\ref{eq:Knunu}), we introduce a tuning parameter $\zeta$.
When $\zeta = 1$ and $0$, the contribution of transverse field is considered fully and absent, respectively.
}

\section{Spectra of induced current density}
\label{sec:excitation:spectra}

{In this section, we examine the influence of {the} T field on the coherent coupling between the individual and collective excirations.
The effect of T field appears as the changes in the hight and shift of the spectral peaks. 
In the following examination, we show the shift of the spectral peaks of collective excitations.
The results are consistent with the shift shown in Ref.\ \cite{iio1}
which is obtained from the evaluation of self-sustained modes obtained from Det$[\bar{\Xi} (\omega)] = 0$,
where $\bar{\Xi} (\omega)$ is the coefficient matrix in Eq.\ (\ref{eq:exSCmat}).
Further, the modulation by T field should appear in the spectral region around the individual excitations
through the T field-mediated shift of the collective excitation.
We see this effect in the modulated spectra of the individual and collective excitations in the following analyses.
Although the T field effect in the present model is not very remarkable because of the small set of model electronic system,
the results should give insight into the possible effect in realistic systems.}

\subsection{Overview of the excitation spectra}


{In this study, we demonstrate the spectrum of collective excitation of electrons in elongated nanorods along the $z$-direction.
In such nanostructures, the collective excitation modes and the induced charge density distribution show one-dimensional-like behavior.
The plasmon excitation depends strongly on its dimensionality~\cite{FriesenBergersen1980,SantoyoMussot1993}.
For the three-dimensional case, the plasmon dispersion at $|\bm{q}| \approx 0$ is finite and constant for $\bm{q}$.
On the other hand, for the two- and one-dimensional case, the plasmon dispersion is propotinal to $|\bm{q}|^{\frac{1}{2}}$ and $|\bm{q}|$,
respectively, hence the excitation starts with zero frequency.
In our demonstration for the nanorods, the spectra follow the one-dimensional-like behavior.}

\begin{figure}
\begin{tabular}{cc}
\includegraphics[keepaspectratio, scale=0.45]{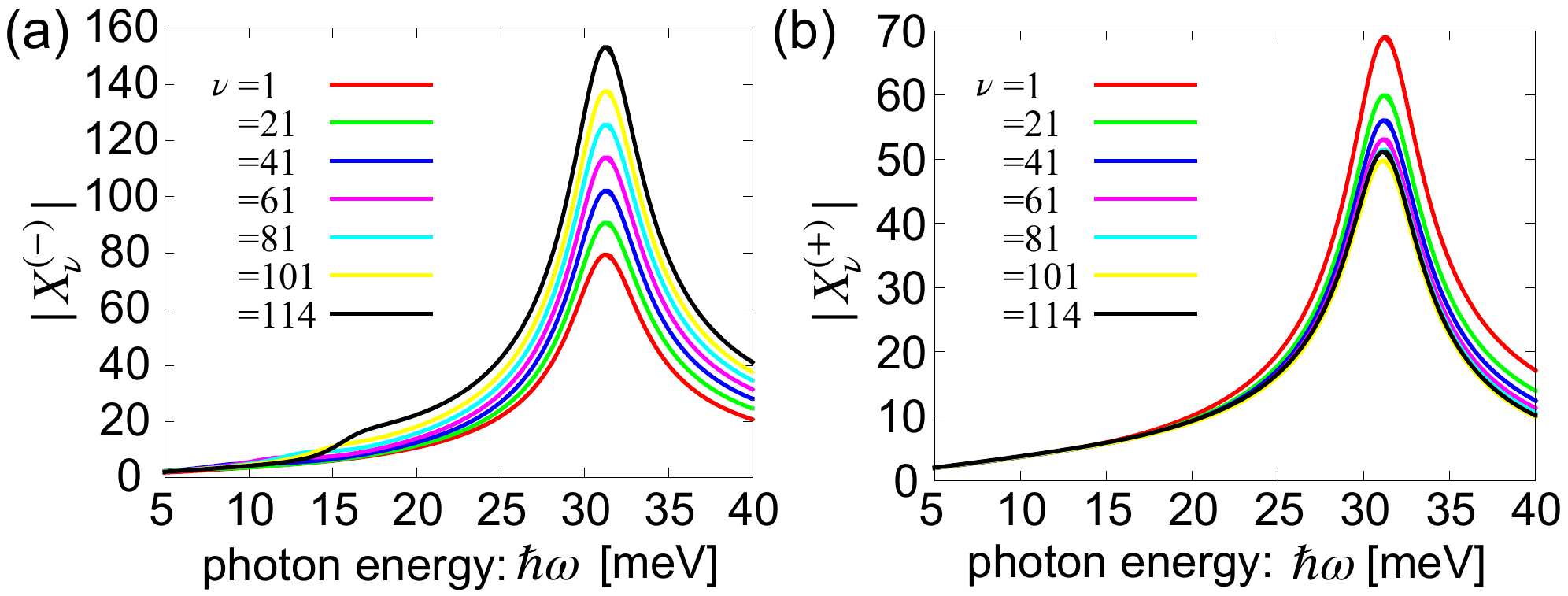}
\end{tabular}
\caption{
Excitation spectra of (a) {the} resonant term {$X_{\nu}^{(-)} (\omega)$} and
(b) {the} non-resonant term {$X_{\nu}^{(+)} (\omega)$ when the T field-mediated interaction is fully considered ($\zeta = 1$).}
{The} ffective mass is fixed as $m^* = 0.07 m_{\rm {e}}$,
{the} Fermi energy {is} $\varepsilon_{\rm {F}} = 3\, \mathrm{eV}$,
{the} background refractive index {is} $n_{\rm b} = 5$, and
{the} nanorod size {is $L_x = 10\, \mathrm{nm}$, $L_y = 15\, \mathrm{nm}$, and} $L_{{z}} = 500\, \mathrm{nm}$.
{The} individual excitation{s are distributed} at $\hbar \omega \le16\, \mathrm{meV}$ and
{the} collective excitation {is} at $\hbar \omega \approx 31\, \mathrm{meV}$.
 }
\label{fig:overall}
\end{figure}

{Figure} \ref{fig:overall} shows the excitation spectra in the presence of the T field
{when the Fermi energy is $\varepsilon_{\rm F} = 3\, \mathrm{eV}$ and
the nanorod length in the $z$-direction is $L_{{z}} = 500\, \mathrm{nm}$ as one typical example (see also Fig.\ 1 in Ref.\ \cite{iio1}).
Here, we set the background refractive index being $n_{\rm b} = 5$.
For the smallest wavenumber $|\bm{q}| = L_z/\pi$,
we should consider $N=114$ bases of the electron-hole pair state $| \nu \rangle$ ($\nu = 1,2,\cdots ,N$).
From Det[$\bar{\Xi} (\omega)]=0$, we obtain 114 self-sustained modes.
As individual excitations, 113 modes are distributed continuously at $2.2\, \mathrm{meV} \le \hbar \omega \le 15.8\, \mathrm{meV}$
and a single isolated mode is at $\hbar \omega \approx 31\, \mathrm{meV}$ as collective excitation.
As the excitation spectrum, we discuss $|X_{\nu}^{(\pm)} (\omega)|$ evaluated by solving the matrix equation (\ref{eq:exSCmat2}).
Here, $|\nu \rangle$ includes the spin degrees of freedom, hence the spectra $|X_{\nu}^{(\pm)}|$ are doubly degenerate.
Note that each spectrum of $|X_{\nu}^{(\pm)} (\omega)|$ includes the contributions from all excitation modes.
The collective excitation provides a dominant contribution, hence all the spectra indicate peak structures at
$\hbar \omega \approx 31\, \mathrm{meV}$.
The scale of peak width is a few $\mathrm{meV}$, which corresponds to the imaginary part of eigenvalues of the matrix $\bar{\Xi}$.}

{Figures \ref{fig:overall}(a) and (b) show $|X_{\nu}^{(-)} (\omega)|$ and $|X_{\nu}^{(+)} (\omega)|$, respectively.
In the definition of $X_{\nu}^{(\pm)} (\omega)$ in Eqs.\ (\ref{eq:X-JA}) and (\ref{eq:X+JA}),}
$\hbar \omega_{{\nu} 0}$ is {the} energy of electron-hole pair of state {$|\nu =(\rm{e}\nu ,\rm{h}\bar{\nu}) \rangle$}
and $\hbar \omega$ is the incident light energy.
{Because of $\hbar \omega_{\nu0} - \hbar \omega$ in the denominator, $X_{\nu}^{(-)} (\omega)$ means} the resonant term.
{In Fig.\ \ref{fig:overall}(a), the spectra indicate shoulder structures at the energy of individual excitations.
On the other hand, $X_{\nu}^{(+)} (\omega)$ is the anti-resonant term,
where the individual excitations are not significant in Fig.\ \ref{fig:overall}(b).}
{Both $X_{\nu}^{(-)}$ and $X_{\nu}^{(+)}$ show main peaks by the collective excitations at $\hbar \omega \approx 31\, \mathrm{meV}$.
For the individual excitations distributed at $2.2\, \mathrm{meV} \le \hbar \omega \le 15.8\, \mathrm{meV}$,
$X_{\nu}^{(-)}$ indicates shoulder structure. However, $X_{\nu}^{(+)}$ does not have.
Moreover, the peak height of $X_{\nu}^{(-)}$ is doubly larger than that of $X_{\nu}^{(+)}$.
This is due to an enhancement by the factor $\hbar \omega_{\nu0} \mp \hbar \omega$.
Hence, the peak height for larger $\nu$ is higher for $X_{\nu}^{(-)}$ and is lower for $X_{\nu}^{(+)}$.}

When the Fermi energy is $\varepsilon_{\rm {F}} = 3\, \mathrm{eV}$ in the present system,
{$N=114$ bases of the electron-hole pair states should be considered for $|\bm{q}| = \pi/L_z$.
For the states of the highest energy ($\nu = 113$ and $114$), the spectra of $X_{\nu = 113,114}^{(-)}$,
which are doubly degenerate, have higher peak than those of the other $X_\nu^{(-)}$ at
the collective excitation ($\varepsilon_{\rm c} \approx 31\, \mathrm{meV}$) and
shows clearly a shoulder structure around the highest individual excitation ($\varepsilon_{\rm i, highest} \approx 16\, \mathrm{meV}$).
Thus, we focus on $X_{\nu = N}^{(-)}$ in the followings.
The coherent coupling between the collective and individual excitations might depend on the energy distance of them,
which can be tuned by the nanorod length $L_z$, the refractive index $n_{\rm b}$, and the Fermi energy $\varepsilon_{\rm {F}}$.
Then, we discuss the $L_z$-, $n_{\rm b}$-, and $\varepsilon_{\rm F}$-dependences} of the resonant spectra
to investigate the relationship between {the} individual and collective excitations.

\subsection{$L_{{z}}$-dependence}

\begin{figure}
\begin{tabular}{cc}
\includegraphics[keepaspectratio, scale=0.45]{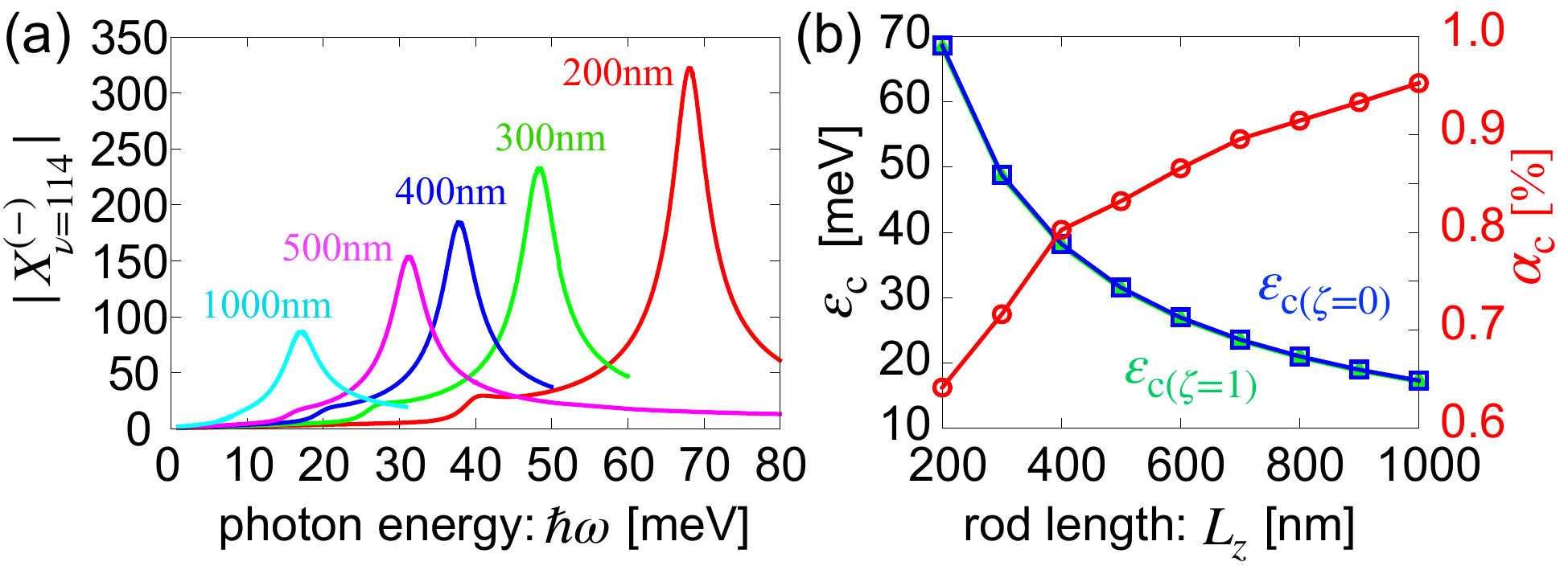}
\end{tabular}
\caption{
{(a) Excitation spectra {$|X_{\nu = 113,114}^{(-)} (\omega)|$} with the T field {($\zeta =1$) for various $L_z$}
when $n_{\rm b}=5$ {and $\varepsilon_{\rm {F}} = 3\, \mathrm{eV}$ are fixed.
The other parameters are the same as those in Fig.\ \ref{fig:overall}.}
(b) {Peak position $\varepsilon_{{\rm c}(\zeta = 1,0)}$ of $|X_{\nu = 113,114}^{(-)} (\omega)|$ for} each $L_z$
{when} $\zeta=1$ (green marks) and $\zeta=0$ (blue marks) {by left axis}
and the modulation ratio
{$\alpha_{\rm c} = (\varepsilon_{{\rm c},(\zeta=0)} - \varepsilon_{{\rm c},(\zeta=1)})/\varepsilon_{{\rm c},(\zeta=0)} \times 100\%$}
of these energies (red marks) {by right axis.
The parameter $\zeta$ tunes the contribution of the T field.}
}}
\label{fig:collective;Lz}
\end{figure}

First, we investigate $L_{{z}}$-dependence {when $n_{\rm b}=5$ and $\varepsilon_{\rm F} = 3\, \mathrm{eV}$ are fixed.}
Figure \ref{fig:collective;Lz}(a) shows spectra of $|X_{{\nu=N}}^{(-)}|$ for {various} $L_{{z}}$.
{Note that even for the change of $L_z$, the number of bases is fixed if the other system parameters are fixed.
In the present case, the number of bases is $N=114$  includeing the spin degrees of freedom.
As menthioned in the previous section, the main peak of the spectrum of $|X_\nu^{(-)}|$ is attributed to the collective excitation.
When $L_z = 200\, \mathrm{nm}$, the collective excitation is $\varepsilon_{\rm c} \approx 69\, \mathrm{meV}$ and
the highest individual excitation is $\varepsilon_{\rm i, highest} \approx 40\, \mathrm{meV}$.
Their distance is larger than the peak width due to the collective excitation.
Then, the shoulder structure due to the individual excitation is found clearly.
When $L_z$ is increased, the peak and shoulder structures are shifted to lower energy
according to the shift of the excitations in Fig.\ \ref{fig:collective;Lz}(a).
The energy distance between the collective and individual excitations decreases with the increase of $L_z$.
When $L_z = 1000\, \mathrm{nm}$,
the shoulder structure is not visible clearly.}

{By the decrease of energy distance, the T field-meadiated coupling between the collective and individual excitations should be enhanced.
Then, we examine the tuning of T field by the parameter $\zeta$ for the peak due to the collective excitation.
Figure \ref{fig:collective;Lz}(b) shows a shift of the peak position due to the collective excitation from $\zeta = 1$ to $0$.
From the shift, we define a modulation {ratio} as
\begin{equation}
\alpha_{\rm c} = \frac{\varepsilon_{{\rm c}(\zeta=0)} - \varepsilon_{{\rm c}(\zeta=1)}}{\varepsilon_{{\rm c}(\zeta=0)}}
\times 100 \%,
\end{equation}
where $\varepsilon_{{\rm c}(\zeta=1)}$ and $\varepsilon_{{\rm c}(\zeta=0)}$}
{are the peak energies for the collective excitations (corresponding to the self-sustained modes)
in the presence and absence of the T field, respectively.}
{Although the difference of these peak values between in the presence and absence of T field is 
not clearly visible if comparing the red and green lines in Fig.\ \ref{fig:collective;Lz}(b) in this scale, 
we can see that the modulation {ratio} increases with the increase of $L_{{z}}$ in the blue line.
The modulation {ratio} is about {$\alpha_{\rm c} \approx 0.63 \%$} when $L_{{z}} = 200\, \mathrm{nm}$,
and it increases up to about {$\alpha_{\rm c} \approx 0.95 \%$} when $L_{{z}}=1000\, \mathrm{nm}$.
This modulation is due to the T field-mediated interaction between electron-hole pairs.}

\begin{figure}
\begin{tabular}{cc}
\includegraphics[keepaspectratio, scale=0.6]{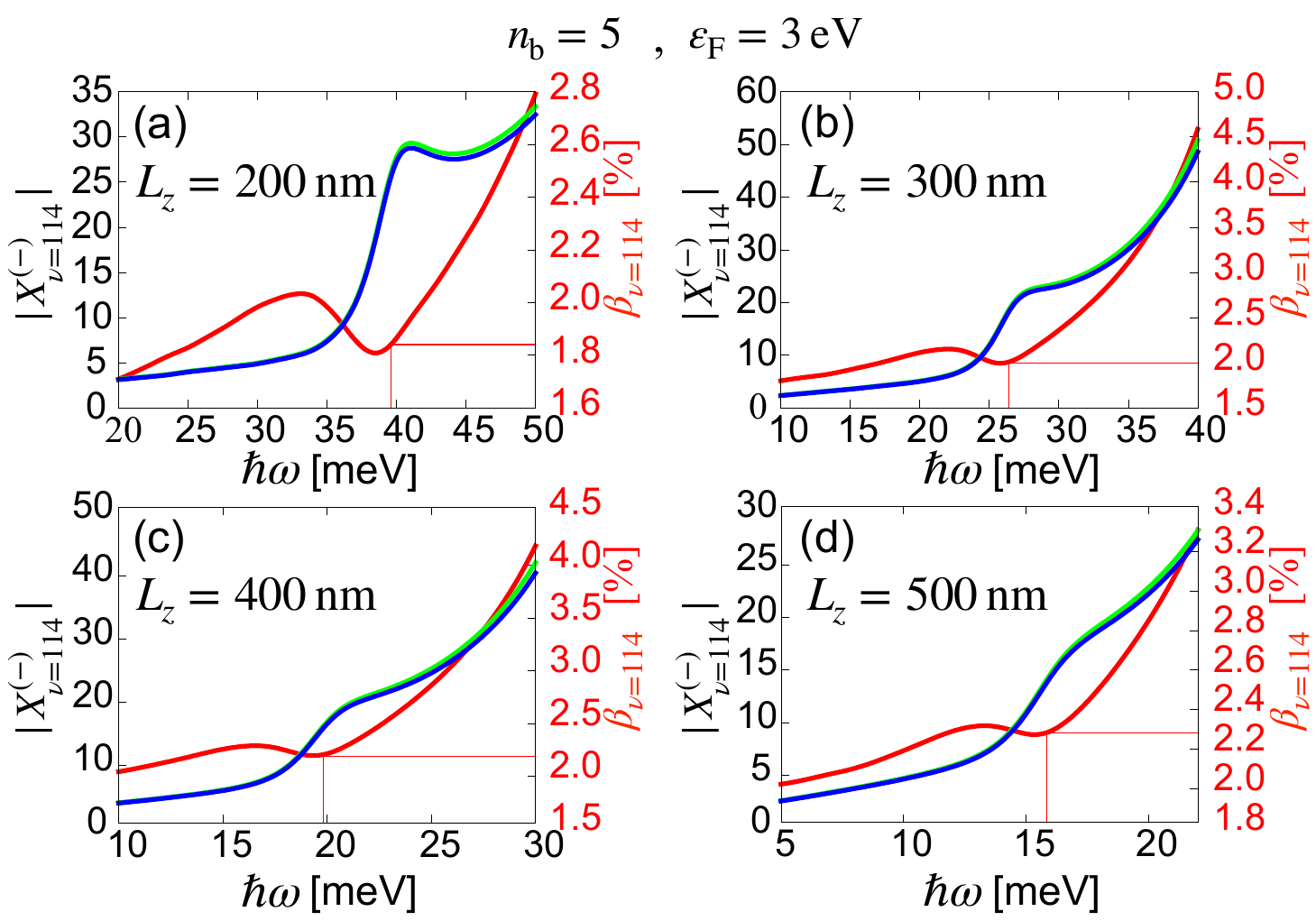}
\end{tabular}
\caption{
{Spectra of $|X_{\nu = N(\zeta = 1,0)}^{(-)}|$} and modulation {ratio $\beta_{\nu = N}$} by the T field
{in the vicinity of} the individual excitation {$\varepsilon_{\rm i,highest}$} for
(a) $L_{{z}}=200\, \mathrm{nm}$, (b) $300\, \mathrm{nm}$, (c) $400\, \mathrm{nm}$, and (d) $500\, \mathrm{nm}$.
{The other parameters are the same as those in Fig.\ \ref{fig:overall}. Then, $N = 114$.
Thick blue and green lines indicate the spectra when $\zeta = 0$ and $\zeta = 1$, respectively, by left axis.
Red line is $\beta_{\nu = N}$ by right axis.}
Thin red line {in each panel indicates $\varepsilon_{\rm i,highest}$} (calculated as a self-sustained mode) and
the modulation {ratio $\beta_{\nu = N}$} at {$\hbar \omega = \varepsilon_{\rm i,highest}$}.
}
\label{fig:norm;nb=5}
\end{figure}

{Although the T field effect appearing in the peak energy shift is not very remarkable for the present small electronic systems,
the effect is more clearly seen {when} we focus on the modulation of {$|X_\nu^{(-)}|$} in the region of individual excitations.}
In Fig. \ref{fig:norm;nb=5}, we demonstrate spectra around the {highest} individual excitation energy {$\varepsilon_{\rm i, highest}$}
with and without the T field.
{In the region near the individual excitations, {it is difficult to discuss the} T field effect as energy shifts of peaks.
However, the modulation of $|X_{{\nu = N}}^{(-)}|$ value appears as the evidence of the coupling
between {the collective and individual} excitations via {the} T field.}

{In the vicinity of the highest individual excitation, the spectrum indicates the shoulder structure.
When we tune the parameter $\zeta$, the spectrum changes slightly.
By this difference between $|X_{\nu (\zeta = 1)}^{(-)}|$ and $|X_{\nu (\zeta = 0)}^{(-)}|$, we evaluate the T field effect.}
{The red lines in Fig.\ \ref{fig:norm;nb=5} show the modulation {ratio} defined as}
{
\begin{equation}
\beta_{\nu} (\omega)=
\frac{|X_{\nu (\zeta=1)}^{(-)} (\omega)| - |X_{\nu (\zeta=0)}^{(-)} (\omega)|}{|X_{\nu (\zeta=0)}^{(-)} (\omega)|}
\times 100 \%.
\end{equation}
} 
{In the modulation {ratio}, we see that the peak-and-dip structure near the individual excitation energy
{$\varepsilon_{\rm i, highest}$} (at the shoulders in the spectra) are visible.}
{Here, the modulation {ratio at} the individual excitation energy {$\varepsilon_{\rm i, highest}$}
increases with the increase of $L_{{z}}$, as shown in Fig.\ \ref{fig:norm;nb=5}(a)-(d).
With the increase of $L_{{z}}$, the individual excitation peaks are buried with the collective excitation,
making it challenging to differentiate between these two types of excitations.}

\begin{figure}
\begin{tabular}{cc}
\includegraphics[keepaspectratio, scale=0.6]{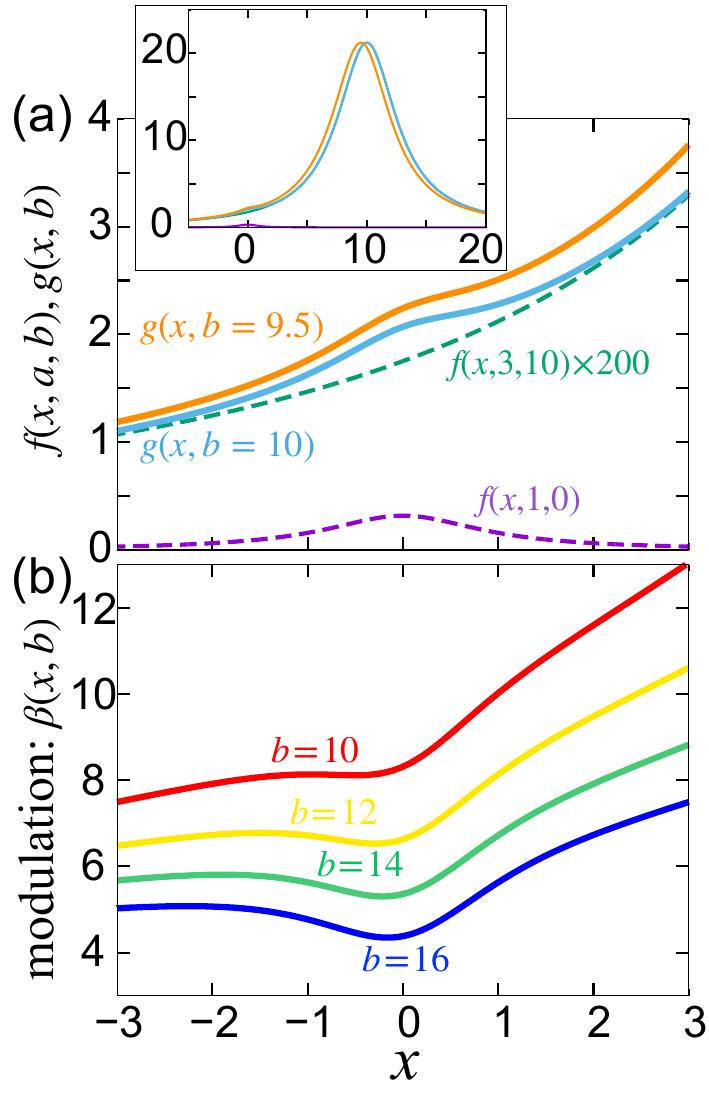}
\end{tabular}
\caption{
Model by two Lorentzian function to explain the behavior of $|X_\nu^{(-)} (\omega)|$ and $\beta_\nu (\omega)$.
(a) Sum of the two Lorentizan function, $g(x,10) = f(x,1,0) + f(x,3,10) \times 200$ (light blue solid line) and
$g(x,9.5) = f(x,1,0) + f(x,3,9.5) \cdot 200$ (orange solid line) with $f(x,a,b) = \frac{1}{\pi} \frac{a}{(x-b)^2 + a^2}$.
Purple and green broken lines are $f(x,1,0)$ and $f(x,3,10) \cdot 200$, respectively.
Inset is a large plot of the functions.
(b) Plot of the ratio $\beta (x,b) = \frac{g(x,b-0.5) - g(x,b)}{g(x,b)} \times 100\%$ for
the shift of large peak position from $x=b$ to $x=b-0.5$.
Red, yellow, green, and blue solid lines represents $\beta (x,b)$ with
the large peak being at $b=10$, $b=12$, $b=14$, and $b=16$, respectively.
}
\label{fig:model}
\end{figure}

{These peak-and-dip behavior of the modulation {ratio $\beta_\nu$ is caused by
the shift of collective excitation $\varepsilon_{\rm c}$ due to the T field.}
Namely, around the individual excitation energy (at the shoulder in the spectra),
the tail of the peak of collective excitation and that of individual excitation are superposed.
{To understand the peak-and-dip behavior by the collective and the (highest) individual excitations,
we examine a superposition of two Lorentizan peaks in Fig.\ \ref{fig:model}.
The Lorentizan peak is given as
\begin{equation}
f(x,a,b) = \frac{1}{\pi} \frac{a}{(x-b)^2 + a^2},
\end{equation}
where $a$ and $b$ means the peak width and position, respectively.
Figure \ref{fig:model}(a) indicates the sum of a small peak at $x=0$, $f(x,1,0)$, and a broad large peak at $x=10$, $f(x,3,10) \times 200$.
They model the peaks by the individual and collective excitations, respectively.
The sum $g(x,10) = f(x,1,0) + f(x,3,10) \times 200$ shows a shoulder structure at $x \approx 0$.
When the large peak shifts slightly to $x=9.5$ from $x=10$ (by the T field), the soulder structure also changes slightly.
For this chage, we calculate $\beta(x,b) = \frac{g(x,b-0.5) - g(x,b)}{g(x,b)} \times 100\%$ as a modulation ratio in Fig.\ \ref{fig:model}(b).
The modulation ratio exhibits a peak-and-dip structure at $x<0$, which is lower than the small peak (individual excitation)~\cite{com2}.
When the position of large peak is changed from $x=10$ to $x=16$,
the dip structure becoms sharp slightly although the ratio is suppressed.
This behabior agrees with $\beta_{\nu = 114}$ by changing $L_z$ in Fig.\ \ref{fig:norm;nb=5}.
When $L_z$ is shorter, the distance between the individual and collective excitations is longer,
then the modulation ratio is smaller.
}
In this way, we understand that the T field effect in the coherent coupling between the individual and collective modes
becomes remarkable with the increase of the rod length $L_{{z}}$.}

\subsection{$n_{\rm b}$-dependence}

\begin{figure}
\begin{tabular}{cc}
\includegraphics[keepaspectratio, scale=0.45]{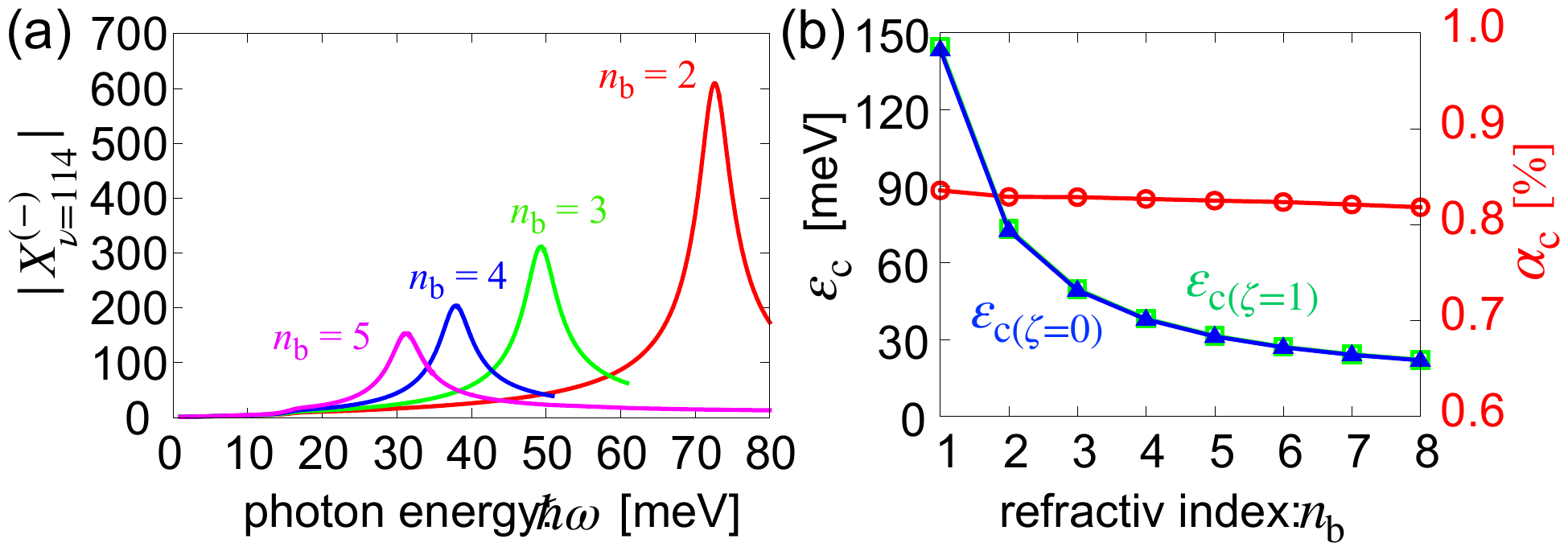}
\end{tabular}
\caption{
{
(a) Excitation spectra $|X_{\nu=114}^{(-)} (\omega)|$ with the T field ($\zeta =1$) for various $n_{\rm b}$
when $L_z=500\, \mathrm{nm}$ and $\varepsilon_{\rm F} = 3\, \mathrm{eV}$ are fixed.
(b) Peak position $\varepsilon_{{\rm c}(\zeta = 1,0)}$ of $|X_{\nu=114}^{(-)} (\omega)|$ for each $n_{\rm b}$
when $\zeta=1$ (green marks) and $\zeta=0$ (blue marks) by left axis and the modulation ratio
$\alpha_{\rm c} = (\varepsilon_{{\rm c},(\zeta=0)} - \varepsilon_{{\rm c},(\zeta=1)})/\varepsilon_{{\rm c},(\zeta=0)} \times 100\%$ of
these energies (red marks) by right axis.
The parameter $\zeta$ tunes the contribution of the T field.}
}
\label{fig:collective;nbg}
\end{figure}

{Figure \ref{fig:collective;nbg} shows the $n_{\rm b}$-dependence of spectra
when {the nanorod length} is fixed as $L_{{z}}=500\, \mathrm{nm}$.
{The Fermi energy is $\varepsilon_{\rm F} = 3\, \mathrm{eV}$.
For this nanorod, the number of bases is $N = 114$ and we examine $X_{\nu =114}^{(-)} (\omega)$.
In} Fig.\ \ref{fig:collective;nbg}{(a)}, we find that {the} collective excitation peaks become smaller with the increase of $n_{\rm b}$.
The strengths of collective excitation decrease monotonically by $1/n_{\rm b}^2$ with the increase of $n_{\rm b}$
(shown in Fig.\ \ref{fig:collective;nbg}{(b)}).
This is because the screening effect by the background refractive index $n_{\rm b}$,
where the Coulomb interaction between electron-hole pairs becomes weak and energy of plasmons is reduced~{\cite{Yokoyama22,iio1}}.
Further, the collective excitation becomes close to the distributed region of individual excitations when $n_{\rm b}$ increases.
Then, it becomes indistinguishable from the individual excitations at larger $n_{\rm b}$,
which means that the collective mode could not be formed due to weak Coulomb interaction at large $n_{\rm b}$.}

{Regarding the modulation {ratio $\alpha_{\rm c}$ due to the T field,
which is evaluated by the peak position of the collective excitation in Fig.\ \ref{fig:collective;nbg}(b)}, 
no remarkable change with {the increase of} $n_{\rm b}$ can be seen.
This is because the reduction of the peak energy due to {the} Coulomb screening compensates the effect of the shortening of
corresponding wavelength by the increase of $n_{\rm b}$.
However, we should note that the coupling between the individual and collective excitations becomes remarkable when $n_{\rm b}$ increases.}

\begin{figure}
\begin{tabular}{cc}
\includegraphics[keepaspectratio, scale=0.6]{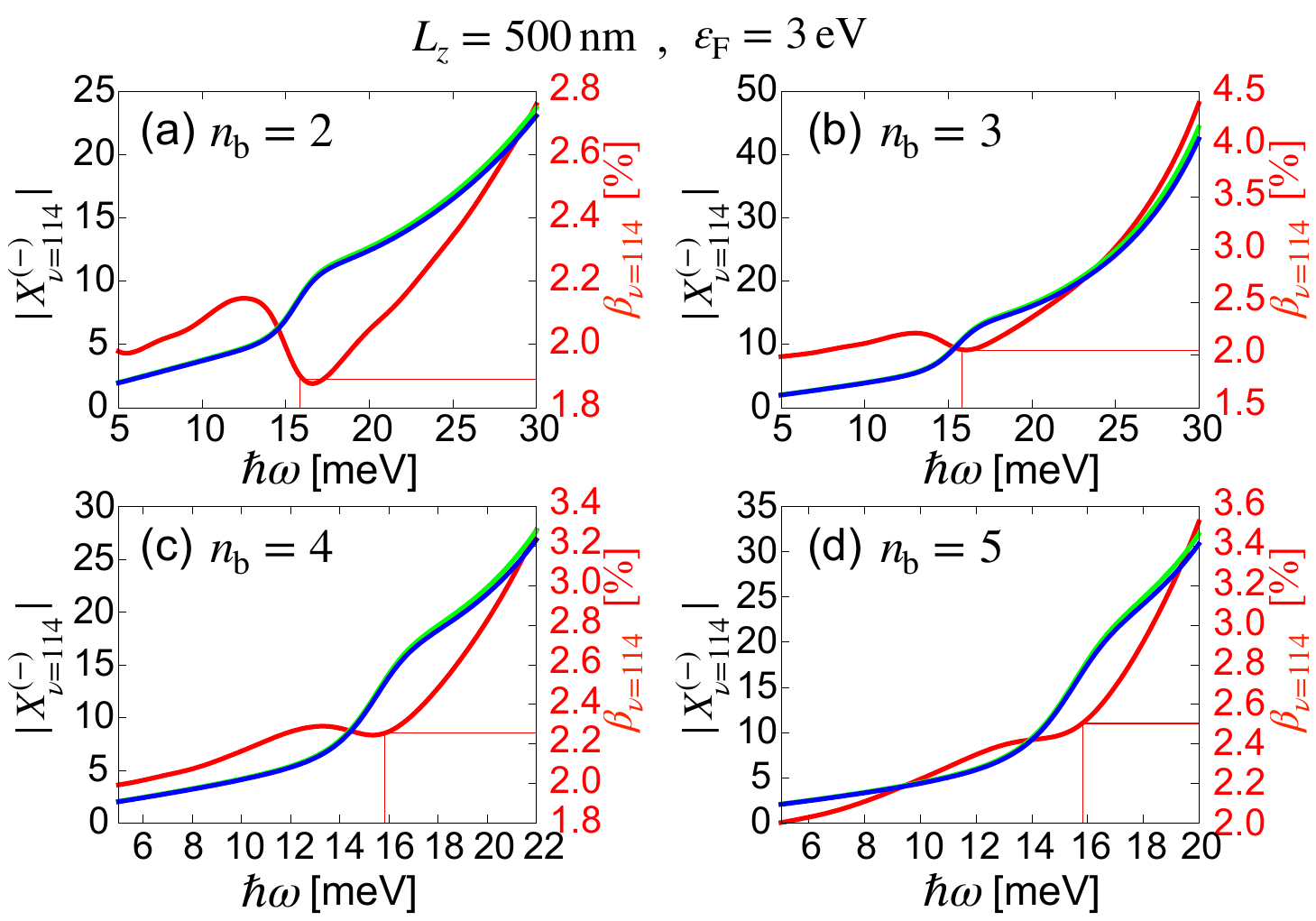}
\end{tabular}
\caption{
{Spectra of $|X_{\nu=114 (\zeta = 1,0)}^{(-)}|$ and modulation ratio $\beta_{\nu=114}$ by the T field
in the vicinity of the individual excitation $\varepsilon_{\rm i,highest}$ for (a) $n_{\rm b}=2$, (b) $3$, (c) $4$, and (d) $5$.
The other parameters are the same as those in Fig.\ \ref{fig:overall}.
Thick blue and green lines indicate the spectra when $\zeta = 0$ and $\zeta = 1$, respectively, by left axis.
Red line is $\beta_{\nu=114}$ by right axis.
Thin red line in each panel indicates $\varepsilon_{\rm i,highest}$ (calculated as a self-sustained mode) and
the modulation ratio $\beta_{\nu = 114}$ at $\hbar \omega = \varepsilon_{\rm i,highest}$.}
}
\label{fig:fitting;nb=1-20;Lz=1000nm}
\end{figure}

{Figure {\ref{fig:fitting;nb=1-20;Lz=1000nm}} shows the spectra {$|X_{\nu =114}^{(-)}|$}
around {the} individual excitation and the modulation {ratio $\beta_{\nu =114}$} by {the} T field
when $L_{{z}}=500\, \mathrm{nm}$ {and $\varepsilon_{\rm F} = 3\, \mathrm{eV}$}.
As in Fig.\ \ref{fig:norm;nb=5}, the peak-and-dip structure of the modulation {ratio is found} for each $n_{\rm b}$.
{In addition, $\beta_{\nu =114}$ increases} with the increase of $n_{\rm b}$.}
{Therefore, the background refractive index $n_{\rm b}$ plays the same role as the nanorod length $L_z$,
and the T field-mediated coupling between the individual and collective excitations is enhanced by $n_{\rm b}$
though it does not contribute to the increase of the modulation {ratio $\alpha_{\rm c}$ at} the collective excitation.}

\subsection{{$\varepsilon_{\rm F}$-}dependence}

\begin{figure}
\begin{tabular}{cc}
\includegraphics[keepaspectratio, scale=0.45]{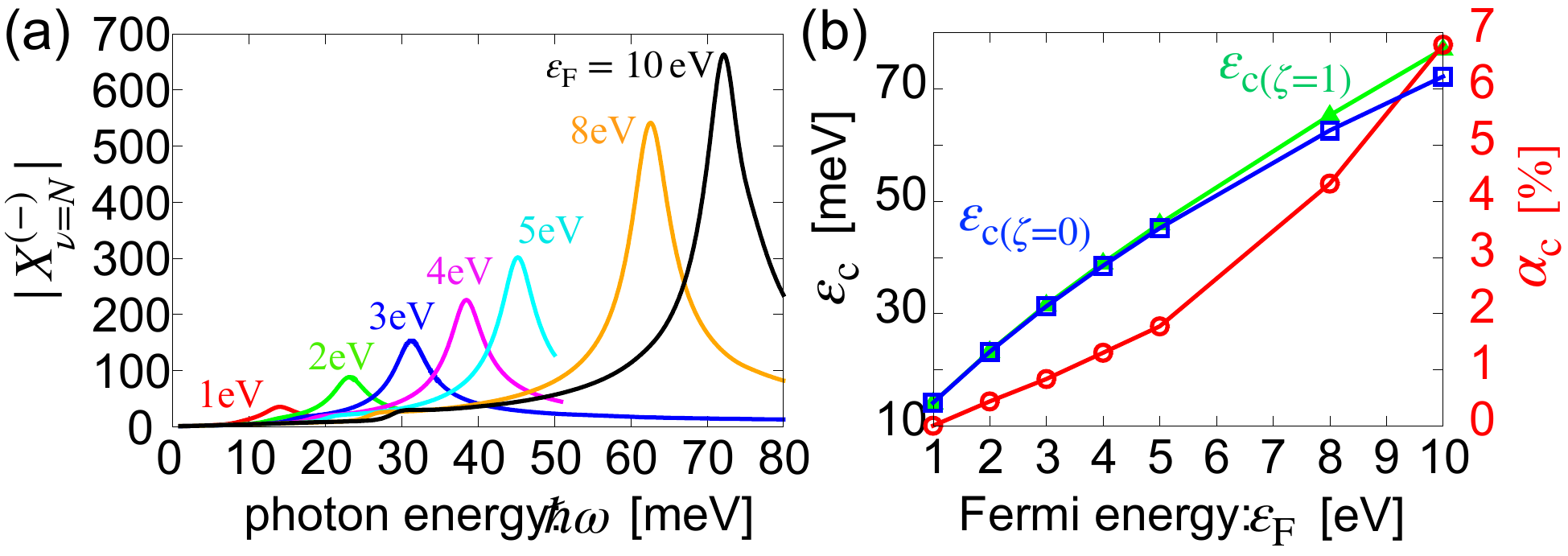}
\end{tabular}
\caption{
{(a) Excitation spectra $|X_{\nu = N}^{(-)} (\omega)|$ with the T field ($\zeta =1$) when the Fermi energy is tuned.
The nanorod length and the background refractive index are fixed at $L_z = 500\, \mathrm{nm}$ and $n_{\rm b}=5$, respectively.
The other parameters are the same as those in Fig.\ \ref{fig:overall}.
When the Fermi energy is $\varepsilon_{\rm F} = 1$, $2$, $3$, $4$, $5$, $8$, and $10\, \mathrm{eV}$,
the number of bases for the electron-hole pair states is $N=34$, $72$, $114$, $154$, $198$, $320$, and $406$, respectively. 
(b) Peak position $\varepsilon_{{\rm c}(\zeta = 1,0)}$ of $|X_{\nu = N}^{(-)} (\omega)|$
when $\zeta=1$ (green marks) and $\zeta=0$ (blue marks) by left axis and the modulation ratio
$\alpha_{\rm c} = (\varepsilon_{{\rm c},(\zeta=0)} - \varepsilon_{{\rm c},(\zeta=1)})/\varepsilon_{{\rm c},(\zeta=0)} \times 100\%$ by right axis.}
}
\label{fig:abs;Ef=3-10eV;Lz=500nm}
\end{figure}

{Finally,} {we demonstrate the Fermi energy dependence of spectrum.
The collective excitation energy increases with the increase of {$\varepsilon_{\rm F}$,
hence the peak in $|X_{\nu = N}^{(-)} (\omega)|$ shifts to higher energy} as shown in Fig.\ \ref{fig:abs;Ef=3-10eV;Lz=500nm}(a).
{Note that the number of bases increase from $N = 34$ at $\varepsilon_{\rm F} =1\, \mathrm{eV}$ to
$N = 406$ at $\varepsilon_{\rm F} =10\, \mathrm{eV}$.
The energy distance between the collective and the highest individual excitations also increases.
When we change $L_z$ and $n_{\rm b}$, the increase of the energy distance results in the decrease of the modulation ratio $\alpha_{\rm c}$.
However, for the tuning of $\varepsilon_{\rm F}$,}
the shift of collective excitation and the modulation {ratio} by the T field increase
{with the increase of energy distance by $\varepsilon_{\rm F}$} in Fig.\ \ref{fig:abs;Ef=3-10eV;Lz=500nm}(b).
The electronic system with the larger {$\varepsilon_{\rm F}$} includes more electrons, and hence, 
these results indicate the possibility that the realistic systems with larger electronic systems,
{the} T field induced change of {the} collective excitations would become significant.}

\begin{figure}
\begin{tabular}{cc}
\includegraphics[keepaspectratio, scale=0.6]{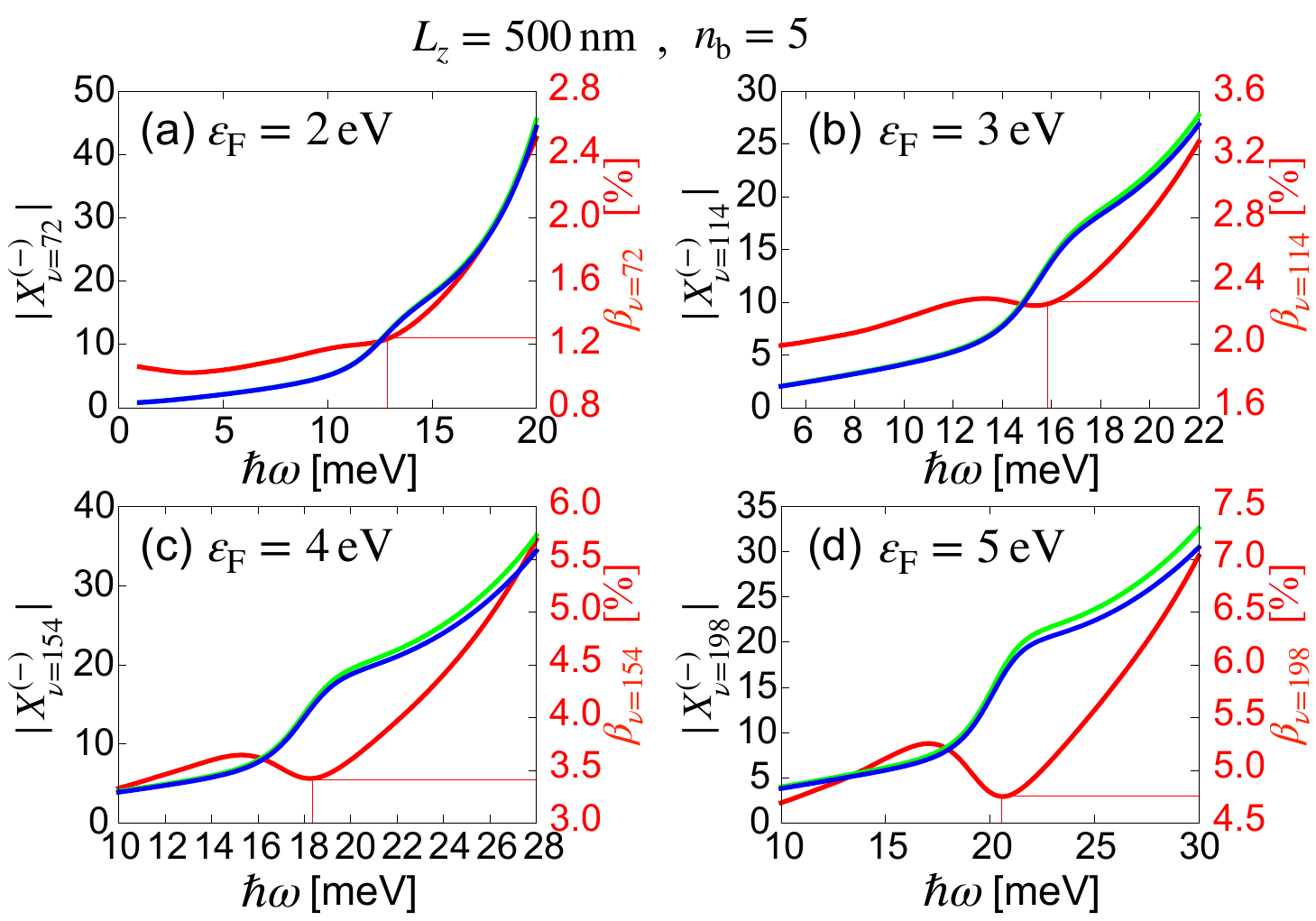}
\end{tabular}
\caption{
{Spectra of $|X_{\nu = N (\zeta = 1,0)}^{(-)}|$ and modulation ratio $\beta_{\nu = N}$ in the vicinity of
the highest individual excitation $\varepsilon_{\rm i,highest}$ when
(a) ${\varepsilon_{\rm F}}=2\, \mathrm{eV}$, (b) $3\, \mathrm{eV}$, (c) $4\, \mathrm{eV}$, and $5\, \mathrm{eV}$.
The other parameters are the same as those in Fig.\ \ref{fig:overall}.
The number of bases is (a) $N=72$, (b) $114$, (c) $154$, and (d) $198$, respectively. 
Thick blue and green lines indicate the spectra when $\zeta = 0$ and $\zeta = 1$, respectively, by left axis.
Red line is $\beta_{\nu = N} (\omega)$ by right axis.
Thin red line in each panel indicates $\varepsilon_{\rm i,highest}$ (calculated as a self-sustained mode) and
the modulation ratio $\beta_{\nu = N}$ at $\hbar \omega = \varepsilon_{\rm i,highest}$.}
}
\label{fig:individual;Ef=3-10eV;Lz=500nm}
\end{figure}

{In Fig.\ \ref{fig:individual;Ef=3-10eV;Lz=500nm}, the spectra $|X_{\nu = N (\zeta)}^{(-)}|$ show
the increase of the modulation ratio $\beta_{\nu = N}$ at the highest individual excitation with the increase of $\varepsilon_{\rm F}$.}
{Around the individual excitation region, the peak-and dip structure becomes remarkable.}

{The Fermi energy can be tuned by doping of carriers in semiconductors or attaching metallic structures.
The increase of $\alpha_{\rm c}$ in Fig.\ \ref{fig:abs;Ef=3-10eV;Lz=500nm}(b) and
$\beta_{\nu = N}$ in Fig.\ \ref{fig:individual;Ef=3-10eV;Lz=500nm} by $\varepsilon_{\rm F}$ is more significant than
those by $L_z$ and $n_{\rm b}$.
Moreover, the increase of modulation ratio by the Fermi energy is opposite to that by
the length (or size) and the refractive index with respect to the energy distance between the collective and the highest individual excitations.
Hence, by a combination of proper tunings of the system paraneters $L_z$, $n_{\rm b}$, and $\varepsilon_{\rm F}$,
the T field-mediated coupling between the individual and collective excitations would be essential in the realistic scale of electronic systems.}

\section{Summary and conclusions}
\label{sec:conculusion}
{Based on the microscopic nonlocal theory, we have investigated the contribution of the transverse radiation field ({transverse} field) 
to the coherent coupling between the collective and individual excitations in metallic materials.
Considering rectangular nanorod model,
we have examined {the transverse} field effect appearing in {the} excitation spectra of induced polarizations.
We have obtained following results: 
The peak energy of collective excitation is shifted by {the transverse} field,
and {it is effectively enhanced} with the increase of the rod length $L_{{z}}$.
{We have evaluated the transverse field effect as the modulation ratio of $|X^{(-)}_{\nu}|$,
which is the component of induced polarization with the highest electronic state.
In addition to the shift of collective excitation, the transverse field causes a peak-and-dip structure of the modulation}
around the individual excitation, {and it is enlarged with the increase of $L_z$.
The transverse} field-induced modulation {exhibits such increaseing} behavior also with
the increase of background refractive index $n_{\rm b}$.}

{
In the present demonstrations, appearance of the {transverse} field effect is not very remarkable
because the model electronic system is small.
However, observing the $L_{{z}}$-dependence of the effect,
we can deduce stronger {transverse} field effect for the larger realistic metallic structures.
Actually, we have demonstrated the stronger effect for the larger Fermi energy {$\varepsilon_{\rm F}$}
where much more electrons are involved in the optical response.
This result gives a good insight into the possible effect for the samples with realistic size scale.}

{In conclusion, it is possible that, in metallic nanostructures, {the transverse} field contributes to the formation of 
collective excitations, and in particular, it should be noted that coherent coupling occurs
between the individual and collective excitations via {the transverse} field. 
The present results lead to a guiding principle to obtain the large coherent coupling between the collective and individual excitations,
which would enable efficient hot carrier generation through a bidirectional energy transfer between 
the collective and individual excitations.}

\begin{acknowledgment}


{This work was supported in part by JSPS KAKENHI (Grant Number: JP21H05019).}

\end{acknowledgment}


\begin{thebibliography}{9}

\bibitem{Ueno16}
K.\ Ueno, T.\ Oshikiri, and H.\ Misawa,
Chem.\ Phys.\ Chem.\ \textbf{17}, 199 (2016).
\bibitem{Li17}
W.\ Li and J.\ G.\ Valentine,
Nanophotonics \textbf{6}, 177 (2017).
\bibitem{Tatsuma17}
T.\ Tatsuma, H.\ Nishi, and T.\ Ishida,
Chem.\ Sci.\ \textbf{8}, 3325 (2017).
\bibitem{Clavero14}
C.\ Clavero,
Nat.\ Photo.\ \textbf{8}, 95 (2014).

\bibitem{Brongersma15}
M.\ L.\ Brongersma, N.\ J.\ Halas, and P.\ Nordlander,
Nat.\ Nanotech.\ \textbf{10}, 25 (2015).
\bibitem{Besteiro17}
L.\ V.\ Besteiro, X.-T.\ Kong, Z.\ Wang, G.\ Hartland, and A.\ O.\ Govorov,
ACS Photonics \textbf{4}, 2759 (2017).
\bibitem{White12}
T.\ P.\ White and K.\ R.\ Catchpole,
Appl.\ Phys.\ Lett.\ \textbf{101}, 073905 (2012).
\bibitem{Govorov13}
A.\ O.\ Govorov, H.\ Zhang, and Y.\ K.\ Gun'ko,
J.\ Phys.\ Chem.\ C \textbf{117}, 16616 (2013).
\bibitem{Govorov14}
A.\ O.\ Govorov, H.\ Zhang, H.\ V.\ Demir, and Y.\ K.\ Gun'ko,
Nano Today \textbf{9}, 85 (2014).
\bibitem{Kumar19}
P.\ V.\ Kumar, T.\ P.\ Rossi, D.\ Marti-Dafcik, D.\ Reichmuth, M.\ Kuisma,
P.\ Erhart, M.\ J.\ Puska, and D.\ J.\ Norris,
ACS Nano \textbf{13}, 3188 (2019).
\bibitem{Goykhman11}
I.\ Goykhman, B.\ Desiatov, J.\ Khurgin, J.\ Shappir, and U.\ Levy,
Nano Lett.\ \textbf{11}, 2219 (2011).

\bibitem{XShi18}
X.\ Shi, K.\ Ueno, T.\ Oshikiri, Q.\ Sun, K.\ Sasaki, and H.\ Misawa,
Nat.\ Nanotech.\ \textbf{13}, 953 (2018). 
\bibitem{YELiu23}
Y.-E.\ Liu, X.\ Shi, T.\ Yokoyama, S.\ Inoue, Y.\ Sunaba, T.\ Oshikiri, Q.\ Sun, M.\ Tamura, H.\ Ishihara, K.\ Sasaki, H.\ Misawa,
private communications.

\bibitem{Ma15}
J.\ Ma, Z.\ Wang, and L.-W.\ Wang,
Nat.\ Commun.\ \textbf{6}, 10107 (2015).
\bibitem{You18}
X.\ You, S.\ Ramakrishana, and T.\ Seideman,
J.\ Phys.\ Chem.\ Lett.\ \textbf{9}, 141 (2018).

\bibitem{Bennett70}
A.\ J.\ Bennett,
Phys.\ Rev.\ B \textbf{1} 203 (1970).
\bibitem{Schwartz82}
C.\ Schwartz and W.\ L.\ Schaich,
Phys.\ Rev.\ B \textbf{26} 7008 (1982).
\bibitem{Pitarke07}
J.\ M.\ Pitarke, V.\ M.\ Silkin, E.\ V.\ Chulkov, and P.\ M.\ Echenique,
Rep.\ Prog.\ Phys.\ \textbf{70}, 1 (2007).
\bibitem{Mortensen14}
N.\ A.\ Mortensen, S.\ Raza, M.\ Wubs, T.\ S{\o}ndergaard, and S.\ I.\ Bozhevolnyi,
Nat.\ Commun.\ \textbf{5}, 3809 (2014).
\bibitem{Christensen14}
T.\ Christensen, W.\ Yan, S{\o}ren Raza.\ A.-P.\ Jauho, N.\ A.\ Mortensen, and M. Wubs,
ACS Nano \textbf{8}, 1745 (2014).
\bibitem{Svendsen20}
M.\ K.\ Svendsen, C.\ Wolff, A.-P.\ Jauho, N.\ A.\ Mortensen, and C.\ Tserkezis,
J.\ Phys.: Condens.\ Matter \textbf{32}, 395702 (2020), and related reference therein.

\bibitem{Yokoyama22}
T.\ Yokoyama, M.\ Iio, T.\ Kinoshita, T.\ Inaoka, and H.\ Ishihara,
Phys.\ Rev.\ B \textbf{105}, 165408 (2022). 

\bibitem{iio1}
M.\ Iio, T.\ Yokoyama,  T.\ Inaoka, and H.\ Ishihara,
arXiv: 2311. 04460 (2023).

\bibitem{book:cho1}
K.\ Cho, {\it Optical Response of Nanostructures: Microscopic Nonlocal Theory}
Springer Series in Solid-State Sciences (Springer-Verlag, Tokyo) 2003.
\bibitem{Kubo57}
R.\ Kubo,
J.\ Phys.\ Soc.\ Jpn., \textbf{12}, 570 (1957).

\bibitem{com1}
Assumed effective mass of electrons corresponds to the value of the conduction band of GaAs.
However, this assumption is just for a typical value of the effective mass and
our discussion is not restricted for a specific material.


\bibitem{FriesenBergersen1980}
W.\ I.\ Friesen and B.\ Bergersen,
J.\ Phys.\ C: Solid St.\ Phys.\ \textbf{13}, 6627 (1980).

\bibitem{SantoyoMussot1993}
B.\ M.\ Santoyo and M.\ del Castillo-Mussot,
Rev.\ Mex.\ Fisica \textbf{4}, 640 (1993).

\bibitem{com2}
For a model calculation by two Lorentzian peaks, a peak-and-dip structure on the moduration ratio could obtain
if the samll Lorentzian (due to individual excitaion) at $x=0$ is also shifted to $x<0$ slightly.
However, in this case, the peak and dip are located at $x<0$ and $x>0$.
Such behabior of the dip and peak on the modulation ratio does not explain the numerical results in
Figs.\ \ref{fig:norm;nb=5}, \ref{fig:fitting;nb=1-20;Lz=1000nm}, and \ref{fig:individual;Ef=3-10eV;Lz=500nm}.

\end{thebibliography}
\end{document}